\documentclass[aps,pra,reprint]{revtex4-2}
\usepackage[utf8]{inputenc}
\usepackage[english]{babel}
\usepackage[T1]{fontenc}
\usepackage{bm}
\usepackage{amsmath}
\usepackage{amssymb}
\usepackage{graphicx}
\usepackage{xcolor}
\usepackage{tikz}
\usepackage{etoolbox}
\usepackage{algpseudocode} 
\usepackage{algorithm}
\usepackage[normalem]{ulem}
\usepackage{float}
\usepackage{comment}
\usepackage{lipsum}
\usepackage{cancel}
\usepackage{multirow}
\usepackage{adjustbox}
\usepackage{xr}
\usepackage{tabularx}
\usepackage[most]{tcolorbox}
\AtBeginEnvironment{tcolorbox}{\small}
\allowdisplaybreaks

\newcommand{\figref}[1]{Fig.~\ref{fig:#1}}

\newcommand{\figrefbegin}[1]{Figure~\ref{fig:#1}}

\newcommand{\secref}[1]{Sec.~\ref{sec:#1}}

\renewcommand{\eqref}[1]{Eq.~(\ref{eq:#1})}
\addto\captionsenglish{\renewcommand{\figurename}{Fig.}}

\renewcommand{\vec}[1]{\mathbf{#1}}

\newcommand{\KET}[1]{|\!\left.{#1}\right>}

\newcommand{\Scale}[2][4]{\scalebox{#1}{$#2$}}

\usepackage[colorlinks=true,
            linkcolor=blue,
            citecolor=blue,
            urlcolor=blue]{hyperref}
\begin{document}

\title{Gradient-free pulse optimization for adiabatic control\\ in open few-body quantum systems}

\author{D. Turyansky}
\thanks{These authors contributed equally to this work.}
\affiliation{Department of Applied Physics, The Hebrew University of Jerusalem, Jerusalem 9190401, Israel}

\author{Y. Zolti}
\thanks{These authors contributed equally to this work.}
\affiliation{Department of Applied Physics, The Hebrew University of Jerusalem, Jerusalem 9190401, Israel}

\author{Y. Cohen}
\affiliation{Department of Applied Physics, The Hebrew University of Jerusalem, Jerusalem 9190401, Israel}

\author{A. Pick}
\email{adi.pick@mail.huji.il}
\affiliation{Department of Applied Physics, The Hebrew University of Jerusalem, Jerusalem 9190401, Israel}

\begin{abstract} 
We present a robust pulse optimization method for adiabatic population transfer and adiabatic quantum computation. The approach relies on identifying control pulses that keep the evolving quantum system close to  its instantaneous ground state. By combining advanced gradient-free optimization tools with specialized cost functions for adiabatic control, it achieves both efficiency and robustness. To demonstrate its generality, we apply the method to three examples involving both atomic and superconducting qubits. We test different optimization cost functions and discretization bases, showing that the approach outperforms ensemble optimization. Finally, to verify its performance on real quantum hardware, we implement digitized adiabatic qubit control using the optimized pulses on the IBM Quantum cloud.
\end{abstract}

\maketitle

\section{Introduction}
We present  a robust pulse optimization method for  adiabatic  control  protocols, including adiabatic  population transfer (APT)~\cite{vitanov2017stimulated} and  adiabatic quantum computation (AQC)~\cite{farhi2000quantum}, where a  driven quantum  system follows  the instantaneous eigenstate of the  Hamiltonian and reaches a desired target state (in APT) or the solution to a computational problem (in AQC).
While adiabatic protocols are inherently robust to fluctuations in the controlling pulses, their speed is limited by the requirement  that the system must  change slowly compared to its spectral gap~\cite{albash2018adiabatic}.  
Therefore, acceleration of adiabatic   protocols is  a topic of active research~\cite{claeys2019floquet,dann2021inertial,garcia2022highly,harutyunyan2023digital,turyansky2024inertial,vzunkovivc2025variational}.
Motivated by the success of quantum optimal control (QOC)  algorithms in enhancing the performance of gate-based quantum algorithms~\cite{evered2023high,jandura2022time}, we introduce a QOC pulse-optimization method  for  adiabatic protocols. \emph{Our method achieves  optimal performance by minimizing   leakage from the instantaneous ground state during the propagation, i.e., the  ``diabatic error.''}  The resulting pulses  possess both high fidelity---by using   QOC---and robustness---due to adherence to the instantaneous ground state. 
  
 To analyze the performance of our method, we  optimize two APT protocols—rapid adiabatic passage in two-level systems (RAP) and multilevel stimulated Raman adiabatic passage (STIRAP)—which are  used as building blocks in AQC~\cite{ebadi2022quantum}. Our final example is an AQC algorithm for the maximum independent set problem (MIS). While the robustness of our pulses to intensity fluctuations is similar to that of  brute-force ensemble optimization~\cite{goerz2014robustness}, our approach is substantially more efficient.

According to the adiabatic theorem, suppressing diabatic errors requires control pulses to vary slowly compared to the spectral gap~\cite{born1928beweis}. In adiabatic qubit control,  increasing the pulse area---the product of the  Rabi frequency  and pulse duration---increases the spectral gap and, therefore, reduces the error. Specifically, In this work, we seek pulses with an integrated power of $4\pi$~\cite{malekakhlagh2024enhanced}. 
However, practical limitations necessitate advanced pulse-engineering methods to achieve optimal performance under realistically attainable pulse areas. Squared trigonometric, hyperbolic, and smooth polynomial pulses are known to effectively minimize diabatic errors~\cite{moller2007geometric,vitanov2017stimulated}. Other methods for  reducing diabatic errors include   parallel-transported eigenvalues~\cite{guerin2002optimization}, superadiabatic states~\cite{lim1991superadiabatic}, and inertial pulses~\cite{turyansky2024inertial,dann2021inertial}. 
A widely used acceleration technique called shortcut-to-adiabaticity (STA)~\cite{del2013shortcuts} involves engineering counter-diabatic driving forces that  suppress diabatic transitions  during the evolution~\cite{berry2009transitionless}. 
While exact STA can eliminate diabatic errors altogether, its construction requires knowledge of all eigenstates and their derivatives~\cite{kolodrubetz2017geometry}. When the latter are known, STA is extremely powerful~\cite{motzoi2009simple,chen2010shortcut,setiawan2021analytic,setiawan2023fast}. However, in large systems, which are required for AQC, the eigenstates are generally unknown. In such cases, one can either use  approximate STA solutions~\cite{takahashi2024shortcuts,finvzgar2025counterdiabatic,morawetz2025universal} or resort to alternative methods.

In this work, we present an acceleration method for adiabatic protocols with QOC. 
Previous applications of QOC to adiabatic control include searching within the subspace of STA solutions for pulses with minimal power and maximal robustness~\cite{mortensen2018fast}, and the application of the chopped random-basis (CRAB) method for optimizing adiabatic entangling gates~\cite{saffman2020symmetric}. Finally, Ref.~\cite{vcepaite2023counterdiabatic} proposes improving approximate STA with additional control fields. Here, we generalize early work by Brif et al.~\cite{brif2014exploring}, which focused on optimizing qubit control with the gradient ascent pulse engineering method (GRAPE)~\cite{khaneja2005optimal}. \emph{By combining gradient-free, multi-objective optimization methods with efficient simulation tools (\figref{flow_chart}), our approach can optimize the controls of few-body open quantum systems.} We validate our method via classical   simulations of  the optimized pulses~\cite{johansson2012qutip}, as well as quantum execution of the digitized pulses~\cite{barends2016digitized} on the IBM quantum platform~\cite{IBMQuantum}.

\section{QOC of adiabatic control pulses}
Real quantum systems are open since they exchange energy and particles with their environment. 
Assuming a Markovian environment with relaxation rate faster than that of the system, the dynamics of the density matrix, $\rho$,  is governed by the Lindblad master equation:~\cite{carmichael2009open}
\begin{equation}
\frac{d\rho}{dt} = -i\left[H(\bm{v}(t)), \rho\right] + \sum_k \left( \hat{L}_k \rho \hat{L}_k^\dagger - \tfrac{1}{2} \{\hat{L}_k^\dagger \hat{L}_k, \rho\} \right).
\label{eq:Lindblad}
\end{equation}
The terms in the sum correspond to dissipative dynamics, with  operators \(\hat{L}_k\) representing   population relaxation and  dephasing.  
The  Hamiltonian \(H\)  includes drift  and control terms. The vector   $\bm{v}(t)$ represents the  control  amplitudes and frequencies that we wish to engineer.

\begin{figure}[t!]
    \centering
    \includegraphics[width=1\linewidth]{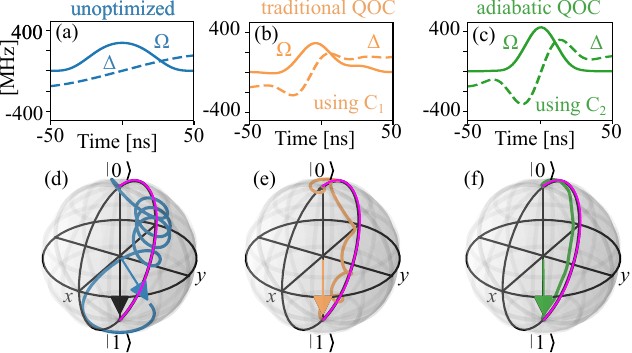}
    \caption{ Three RAP Pulses for adiabatic excitation in a  two-level system  with Rabi frequency $\Omega$ and  detuning $\Delta$: (a) An    unoptimized polynomial pulse, (b) a  pulse obtained with traditional QOC, and  (c) a pulse obtained with adiabatic QOC. 
    (d,e,f) State trajectories on the Bloch sphere  under pulses from (a,b,c). Purple curve is   the  ground-state trajectory. }    \label{fig:RAP-fig1} 
\end{figure}


To accelerate adiabatic control protocols, we seek controls  $\bm{v}(t)$ that drive   the  system along     the instantaneous ground state of $H$. To do so, We minimize  the  functional:
\begin{align}
\mathcal{C}\left(\rho(t),\bm{v}(t),t_f\right) =     \mathcal{C}_i + \eta\,\mathcal{C}_0,
 \label{eq:cost-objective}
\end{align}
where $\eta$ balances infidelity ($\mathcal{C}_i$) and power ($\mathcal{C}_0$) constraints  and $t_f$ is the protocol duration.
The term
$\mathcal{C}_{0} = \sum_{\nu}\int_0^{t_f} v_\nu^2(t)dt - \mathcal{C}_\mathrm{ref}$
reduces total control power relative to the reference $\mathcal{C}_\mathrm{ref}$ (summed over all controls)~\footnote{Here, we use a reference value of $4\pi$~\cite{malekakhlagh2024enhanced}.}.
We consider  three figures of merit for the infidelity:
\begin{subequations}
\begin{align}
C_1 &= 1 - \text{Tr}\textstyle\sqrt{\textstyle\sqrt{\rho_g(t_f,\vec{v}}) \, \rho(t_f,\vec{v}) \, \sqrt{\rho_g(t_f,\vec{v})} },\label{eq:traditional-QOC}\\
C_2 &= \tfrac{1}{t_f}\!\!\int_0^{t_f} \! dt \left(1 - \text{Tr}\textstyle\sqrt{\textstyle\sqrt{\rho_g(t)} \, \rho(t,\vec{v}) \, \sqrt{\rho_g(t)}}\right),\label{eq:adiabatic-QOC}\\
C_3 &= \tfrac{1}{M} \textstyle\sum_{i=1}^M u_i \, C_1\big\{\rho_i(t),\vec{v}(t), \alpha_i)\big\}.\label{eq:ensemble-QOC}
\end{align}
\label{eq:cost-functions}
\end{subequations}
\hspace{-3pt}The term \(C_1\) quantifies the deviation of the system's state at the end of the protocol, \(\rho(t_f)\), from the desired eigenstate, \(\rho_g(t_f)\), using the   fidelity amplitude metric~\footnote{We used the QuTip built-in function to compute the fidelity amplitude. The traditional Uhlmann fidelity~\cite{jozsa1994fidelity} is the square of the QuTip fidelity amplitude.},\footnote{We occasionally omit the dependence of density matrices on the controls vector $\vec{v}(t)$  for ease of notation. }. 
We refer to $C_1$  as the diabatic error, and to pulses obtained by minimizing it as traditional QOC. The cost $C_2$ is the accumulated diabatic error along the propagation, and we refer to pulses obtained by minimizing it as adiabatic QOC.    In the numerical examples that follow, we use a  weighted sum of the infidelity costs, $C_1 + \lambda C_2$, showing  that by minimizing $C_2$, one gains  robustness. The choice of the weight factor $\lambda$ is discussed in in Sec.~1.1 in the supplementary material (SM)~\cite{supplement} and in the following  examples (\secref{Examples}). The cost $C_3$ is used for ensemble optimization~\cite{goerz2014robustness}. It   is defined as the  average diabatic error, $C_1$,
for an ensemble of systems whose states $\rho_i(t)$ evolve under controls    shifted or scaled by $\alpha_i$, with $u_i$ denoting the averaging  weights.

\figrefbegin{RAP-fig1} shows how pulse optimization affects qubit dynamics. We consider the RAP protocol for qubit population transfer (\secref{example1}). In this case, the control fields $\vec{v}(t)$ are the Rabi frequency $\Omega$ and detuning $\Delta$ [\eqref{RAP_hamiltonian}]. We compare three choices of control fields: (a) unoptimized  (Gaussian $\Omega$ and linear $\Delta$), (b) pulses optimized with traditional QOC, and (c) pulses optimized with adiabatic QOC. The bottom plots show the corresponding dynamics on the Bloch sphere, where the purple curve denotes the trajectory of the instantaneous ground state. Evidently,  the system follows the ground state when driven with pulses obtained from adiabatic QOC.


\begin{figure}[b!]
    \centering
    \includegraphics[width=1\linewidth]{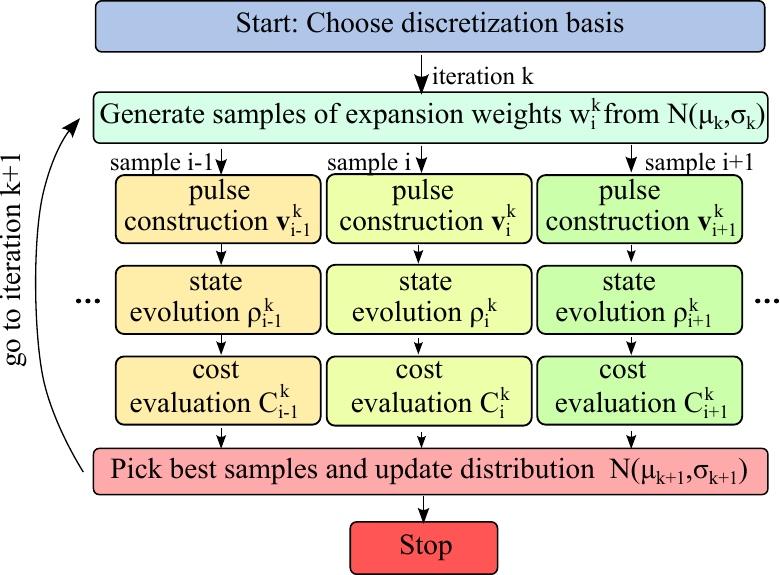}
    \caption{\textbf{Flow chart of  iterative  adiabatic QOC algorithm.} In each iteration ($k$), generate random samples of expansion coefficients ($w_i^k$). For each sample ($i$),  construct pulse, evolve system, and evaluate   cost  using \eqref{cost-functions} (including the adiabatic infidelity cost $C_2$). Select best samples  and update the distribution to lower the cost. }
    \label{fig:flow_chart} 
\end{figure}

An overview of our optimization procedure  is given below. A flow chart is provided in \figref{flow_chart}.  For efficient pulse representation, we write the control pulses as a sum of a reference pulse — which we apriori  know provides good and adiabatic performance — and, on top of it, a correction  written as a finite sum of basis elements whose weights we seek. This is  similar to  the   chopped random basis (CRAB)~\cite{caneva2011chopped}. For optimal performance, we choose between three different  of bases depending on the case of study: 
\begin{subequations}
\begin{align}
v^{(1)}(t) &= v_{0}(t) + \sin^2\left( \tfrac{\pi t}{t_f} \right)\sum_n w_n \Scale[1]{\exp\left(-\tfrac{(t - \mu_n)^2}{\sigma_n^2}\right)}, \label{eq:gaussians}\\
v^{(2)}(t) &= v_{0}(t) + \sin^2\left( \tfrac{\pi t}{t_f} \right)\sum_n w_n \sin\left( \tfrac{n\pi t}{t_f}\right),\label{eq:fourier}\\
v^{(3)}(t) &= v_{0}(t) + \sin^2\left( \tfrac{\pi t}{t_f} \right)  \sum_n w_n T_n\left(\tfrac{2t}{t_f} - 1\right), \label{eq:cheby} 
\end{align}
\label{eq:basis}
\end{subequations}
\hspace{-4pt}where \(v_{0}(t)\) is a fixed reference pulse and the sums  represent corrections spanned in bases of  Gaussians, sine functions, and Chebyshev polynomials, $T_n(z)$, respectively.  The squared sine envelope in \eqref{cheby} guarantees smoothness of the   basis functions at the start and end of the protocol~\cite{turyansky2024inertial}.  
In Sec.~1.2 in  the SM~\cite{supplement}, we compare the  performance across different bases to justify our basis selection. We find that Chebyshev polynomials consistently yield the lowest infidelity, while Gaussian functions provide the fastest convergence. The sinusoidal basis offers a trade-off between convergence rate  and fidelity.

Once choosing an expansion basis, we employ the CMA-ES algorithm~\cite{hansen2006cma} to find optimal expansion coefficients. 
Evolutionary algorithms like CMA-ES are widely used in experimental QOC~\cite{roslund2009accelerated,shir2012quantum,yarkoni2019boosting,ebadi2022quantum} because they handle multi-objective cost functions well~\cite{shir2012quantum}, as also required in our adiabatic QOC setting. We use the open-source Python library \texttt{Nevergrad}~\cite{bennet2021nevergrad} to implement CMA-ES. The algorithm works by drawing random samples of expansion coefficients from a normal distribution and iteratively  updating the distribution. At each iteration, one generates samples of expansion coefficients. For each sample, one constructs the corresponding pulses, evolves the state under \eqref{Lindblad} (using \texttt{QuTiP}~\cite{johansson2012qutip}), and evaluates the cost using  \eqref{cost-objective}. Finally, the algorithm picks  samples with minimal costs to update the mean and covariance of the  distribution and reiterate until a  stop criterion is reached.  The convergence of CMA-ES is analyzed in the SM~in Sec.~1.3~\cite{supplement}. Next, we apply the optimization algorithm just described to three numerical examples.




\section{Numerical examples\label{sec:Examples}}
\subsection{Optimization of adiabatic qubit control.\label{sec:example1}}
First, we study rapid adiabatic passage (RAP), a protocol for population transfer between the ground and excited states of a two-level system using a chirped pulse. The Hamiltonian in the rotating-wave approximation is~\cite{gerry2023introductory}
\begin{equation}
H_{\rm RAP} = \Omega(t)\hat{\sigma}_x + \Delta(t)\hat{\sigma}_z,
\label{eq:RAP_hamiltonian}
\end{equation}
where $\Omega(t)$ is the Rabi frequency and $\Delta(t)$ is the detuning. In RAP, $\Omega$ vanishes at the beginning and end of the protocol, so the Hamiltonian reduces to $\Delta \hat{\sigma}_z$. Because $\Delta$ changes sign during the sweep, the instantaneous ground state evolves from $\KET{0}$ to $\KET{1}$, enabling population transfer by adiabatically following the ground state.

To improve state-transfer fidelity under a fixed pulse area, we apply our  QOC method. We compare qubit dynamics under an unoptimized polynomial pulse (blue)\footnote{The polynomial pulses we use are: $\Omega(t) = A - B \left(\frac{t}{\tau}\right)^4 + C\left(\frac{t}{\tau}\right)^6$ and $\Delta(t) = a\left(\frac{t}{\tau}\right) - b\left(\frac{t}{\tau}\right)^3$, with coefficients chosen to have vanishing slopes and curvatures at the start and end of the protocol, as proposed in~\cite{turyansky2024inertial}.} with dynamics under two optimized pulses (using the sine basis~\eqref{fourier}): a ``traditional pulse'' obtained using $C_1$~[\eqref{traditional-QOC}] (orange), and an ``adiabatic pulse'' obtained using $C_1 + C_2$ for the infidelity cost~[\eqref{adiabatic-QOC}] (green).
 The   pulses  are shown  in \figref{RAP-fig1}(a--c), and corresponding  Bloch-sphere trajectories in \figref{RAP-fig1}(d--f). As discussed conceptually above, only the adiabatic-QOC pulse tracks the instantaneous ground state.

\begin{figure}[t!]
    \centering
    \includegraphics[width=1\linewidth]{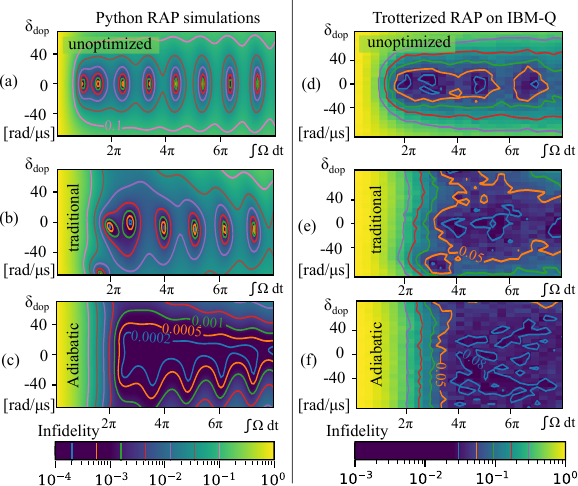}
    \caption{Robustness of optimized  RAP pulses. (a--c) Python simulations showing infidelity of state transfer [\eqref{traditional-QOC}]  (colorbar) for the three pulses from \figref{RAP-fig1} under variations in pulse amplitude, $\int_0^{t_f} \varepsilon \Omega(t) dt$ and frequency shifts, $\Delta(t)+\delta_\mathrm{dopp}$. (d--f) Infidelity of a trotterized RAP protocol under variations in the controls pulse area and detuning, obtained from quantum simulations  on the IBM quantum cloud~\cite{IBMQuantum}.}
    \label{fig:Fig2_robust} 
\end{figure}

Since the adiabatic-QOC trajectory adheres to the instantaneous eigenstate, we expect robustness to fluctuations in the control amplitude and frequency, which are prevalent  in atomic qubits~\cite{kasevich1991atomic}. To assess the robustness under such noise, we evolve the system  under control fields with  a rescaled Rabi frequency, $\Omega(t) \rightarrow \epsilon \Omega(t)$, and  shifted detuning, $\Delta \rightarrow \Delta + \delta_\mathrm{dopp}$ and compute the state-transfer infidelity.
\figrefbegin{Fig2_robust}(a--c) summarizes the results for  the three pulses from \figref{RAP-fig1}, showing the  infidelity over a range of control parameters~\footnote{The $x$ axis denotes the rescaled Rabi frequency in units of pulse area.}. 
Both the unoptimized  pulse and the pulse obtained from traditional QOC  show   high-fidelity islands (dark blue) arising from   Rabi oscillations and a sharp fidelity drop for non-zero Doppler shifts. However, the pulse obtained from adiabatic QOC achieves high state-transfer fidelity over a wider  range of parameters (dark blue). 

To verify our results on real quantum hardware, we ran a trotterized version of the pulses from \figref{RAP-fig1} on  IBM's \texttt{brisbane} quantum computer~\cite{IBMQuantum}~\cite{barends2016digitized}.  We generated 625 Hamiltonian parameter sets, corresponding to 25 Doppler shifts and pulse areas, and then simulated the system evolution and recorded the fidelity for each set. Infidelity averaged  over 1024 runs is shown in \figref{Fig2_robust}(d--f). The measured results resemble   the Python results, but the fidelity of the former is lower due to the limited repetition number at each parameter set. In Sec.~2.1 in the SM~\cite{supplement}, we compare our method with ensemble optimization. While both methods find pulses that are robust to Doppler shifts and intensity variations, adiabatic  optimization is substantially  more efficient.

\begin{figure}[b]
    \centering
    \includegraphics[width=1\linewidth]{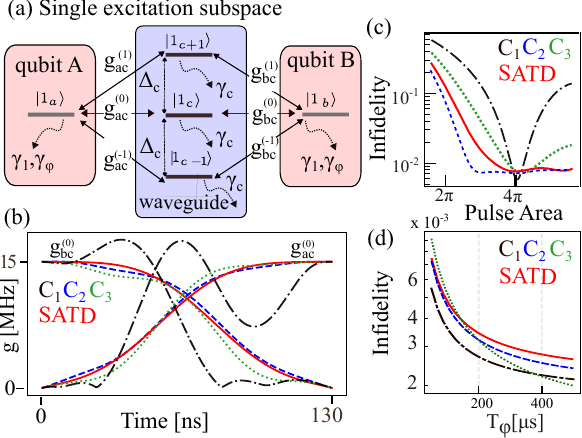}
    \caption{(a) Single-excitation subspace of  two qubits coupled to a multimode waveguide with couplings \(g_{ac}^{(n)}\)  and \(g_{ab}^{(n)}\) and FSR 
    $\Delta_c$. Decay rates  $\gamma_1, \gamma_{\phi}, \gamma_c$ representing qubit relaxation, dephasing, and waveguide decay. 
(b) Coupling amplitudes  \( g_{ac}(t) \)   and \( g_{ab}(t) \)   of  SATD  pulses  and pulses obtained with  $C_1$, $C_1 + \tfrac{1}{2}C_2$ (labeled as $C_2$) and $C_3$ [\eqref{cost-functions}].
(c,d) State-transfer infidelity for varying $T_\varphi = 1/\gamma_\phi$  (where $C_1$ is optimal) and    pulse areas  of the qubit-waveguide couplings (where $C_2$ is optimal) for pulses from (b)
and assuming $Q_c = 10^5$
where  $Q_c = \omega_c / \gamma_c$ and $\omega_c = 5$ GHz and $\Delta_c = 500$ MHz.
}
    \label{fig:stirap-figure}
\end{figure}


\subsection{Adiabatic QOC  of a multilevel system.}

Next, we  accelerate  a protocol  for adiabatic  state transfer between  superconducting qubits mediated    via a multimode waveguide~\cite{chang2020remote}. 
The energy level diagram of the system (reproduced from~\cite{malekakhlagh2024enhanced})  is shown in   \figref{stirap-figure}(a), where qubits $A$ and $B$  couple to  the $n^{th}$  guided mode with amplitudes $g_{ac}^n(t)$ and $g_{bc}^n(t)$ respectively.  
The protocol is a generalized STIRAP scheme~\cite{vitanov2017stimulated}, in which one  first applies    $g_{bc}^n(t)$ and  then $g_{ac}^n(t)$ [\figref{stirap-figure}(b)]. By doing so, the instantaneous ground state changes and excitation is transferred from  $A$ to  $B$. We optimize the pulse shapes,  $g_{bc}^n(t)$ and   $g_{ac}^n(t)$, using our adiabatic QOC algorithm.   To benchmark our method,  we compare it with   super-adiabatic transitionless driving (SATD),  which relies on eliminating diabatic errors in the dressed-state basis~\cite{baksic2016speeding}. The details  of the  SATD pulses are given in Sec.~4 of the SM~\cite{supplement}.  We 
include population relaxation ($T_1$) and dephasing ($T_\varphi$) noise  using the Lindblad master equation [\eqref{Lindblad}].  Its use can be  justified as follows.  While non-Markovian dynamics are generally required for qubits coupled to multimode waveguides~\cite{houck2008controlling, malekakhlagh2016origin, gely2017convergence, malekakhlagh2017cutoff,he2025quantum}, when using   meter-long microwave waveguides, qubit couplings to high-order interconnect modes are nearly frequency independent justifying the Markov approximation underlying \eqref{Lindblad}~\cite{chang2020remote,malekakhlagh2024enhanced}.

\figrefbegin{stirap-figure}(b) shows optimized pulse shapes
\(g_{ac}^{(0)}(t)\)  and \(g_{ab}^{(0)}(t)\) obtained with  SATD  and   QOC using the infidelity costs from \eqref{cost-functions}.   The dashed-blue curve is obtained with adiabatic QOC by  minimizing  the weighted sum $C_1 + \tfrac{1}{2}C_2$.  Here, we use a basis of Chebyshev polynomials since  we prioritize  high fidelity over fast convergence.
\figrefbegin{stirap-figure}(c) and (d) show the robustness of the pulses from (b)  to intensity and lifetime variations respectively. 
In (c),  we compute infidelity  upon   scaling the amplitude of the   pulses from (b) by an overall scale factor, replacing $g_{ac}^{(n)} \rightarrow \epsilon g_{ac}^{(n)}$ and $g_{bc}^{(n)} \rightarrow \epsilon g_{bc}^{(n)}$. We assume  $T_1 = T_\varphi = 100~\mu$s and a waveguide quality factor of $Q =  10^5$.  
In (d) we plot  the state-transfer infidelity versus  $T_\varphi$ for the pulses from (b). The pulses are optimized assuming $T_1 = 50\,\mu\text{s}$. 

Our analysis shows that pulses obtained with adiabatic QOC have improved robustness for intensity fluctuations, but not to variations in   dephasing lifetime. The robustness to amplitude variations can be attributed to the fact that when forcing $g_{bc}^{(0)} = 0$ at $t_f$,  an overall scale factor does not affect the final ground state. Finally, we note that ensemble optimization produces robust results but is significantly less efficient than adiabatic QOC; it takes approximately one hour compared to several minutes with adiabatic QOC.


\subsection{Adiabatic QOC  of a few-body quantum  system.}
\begin{figure}[!htbp]
    \centering
    \includegraphics[width=1\linewidth]{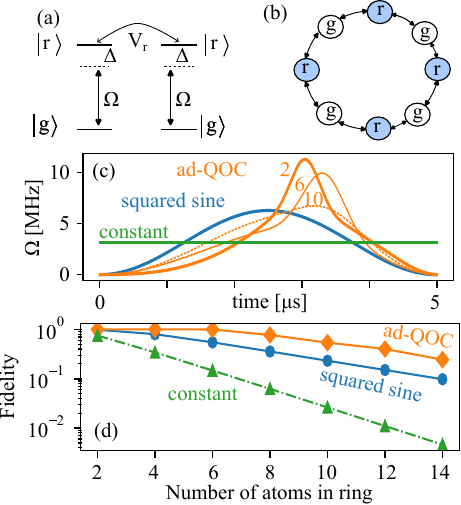}
    \caption{(a) Rydberg qubits with levels  $\KET{g}$ and  $\KET{r}$.  $V_r$ is the Rydberg interaction, while $\Omega$ and $\Delta$ are the Rabi frequency and detuning of the control. (b) An 8-node ring graph  and one of its MIS solutions. (c) Rabi frequency $\Omega(t)$ of a constant field (green), a squared sine pulse (blue), and pulses obtained with adiabatic QOC for rings containing $N = 2, 6, 10$ atoms (orange). 
    (d) Fidelity defined as the overlap of the final state and  MIS solution,  $1- \mathcal{C}_1$, versus ring length for the pulses from (c). }
    \label{fig:MIS-figure} 
\end{figure}

Finally, we accelerate an  AQC algorithm for finding the maximum independent set (MIS) of a graph. Following Refs.~\cite{pichler2018quantum,ebadi2022quantum,kim2024quantum,dalyac2024graph}, we use Rydberg qubits---two-level atoms with a low-energy hyperfine state and a highly excited  Rydberg state [\figref{MIS-figure}(a)]. Rydberg states have a large dipole moment, preventing atoms separated by less than the critical ``blockade radius'' \(r_B\) from being simultaneously excited. A lattice of Rydberg qubits  represents a  unit-disk graph, \(G(N,E)\), where each qubit corresponds to a node \(i \in N\), and nodes represented by qubits with mutual distance \(d < r_B\)  are connected by an edge \((i,j) \in E\). The AQC for finding the MIS of \(G(N,E)\) proceeds as follows~\cite{pichler2018quantum}: All qubits are initialized in the ground state. Then, a global RAP pulse is applied, attempting to excite all atoms.  Rydberg blockade prohibits excitations of qubits with mutual distance \(d < r_B\), so the system naturally finds the maximum number of allowed Rydberg excitations---corresponding to the MIS of \(G(N,E)\). 

The Hamiltonian of the driven Rydberg lattice is
\begin{equation}
H_\mathrm{MIS} = \sum_{i \in N}\left[\Omega(t)\hat{\sigma}_x^i-\Delta(t)\hat{\sigma}_z^i\right]
+V_r \sum_{(i,j)\in E}\hat{n}_i\hat{n}_j,
\end{equation}
where the first sum represents the global RAP pulse, with $\Delta(t)$ a cubic polynomial swept from positive to negative, and $\Omega(t)$ an optimized pulse, shown in \figref{MIS-figure}(c). The second sum represents Rydberg interactions, with $V_r$ the Rydberg interaction strength and $\hat{n}_i$ the projector of qubit $i$ onto the Rydberg state, i.e., the product $\hat{n}_i \hat{n}_j$ is non-zero when both $i$ and $j$ are excited.

We apply our method to study  ring graphs with an even number of nodes. The MIS is known---consisting of every other node [\figref{MIS-figure}(b)]---and is used to evaluate the error of the AQC.  We analyze the  error in finding the MIS when using the pulses shown in  \figref{MIS-figure}(c):  A constant Rabi intensity (green), a squared sine pulse (blue), and pulses obtained from adiabatic QOC for rings containing $2, \dots, 14$ atoms (orange) when using the squared sine pulse as an initial guess. (For clarity, only 3 optimized pulses are shown.) Here, we use a weighted sum of fidelity costs, $C_1+0.1C_2$.  Upon increasing the Hilbert space dimension, we find that convergence improves when decreasing the weight of $C_2$. We demonstrate the robustness of the optimizaed pulses to intensity fluctuations in Sec.~2.2 of the SM~\cite{supplement}. We 
discretize the control pulses with  Gaussian basis  functions since fast convergence is crucial in     multi-qubit control.  To facilitate  convergence for large rings, we reduce the Hilbert space dimension by projecting the Hamiltonian onto the subspace of states  without  neighboring Rydberg excitations~\cite{pichler2018quantum}.  We analyze the  error induced by this projection in Sec.~3 of the SM~\cite{supplement}.  To ensure a fair comparison with the unoptimized pulses in \figref{MIS-figure}(c), we normalize all pulse areas to $4\pi$ after pulse optimization and before running the AQC.  \figref{MIS-figure}(d) shows the fidelity (defined as the overlap between the final  and  true MIS states) versus  chain length for the pulses from (c), demonstrating the superiority of  adiabatic QOC over other methods. For short chains ($N = 2,\dots,8$) we find slightly asymmetric pulses (orange), while for longer chains, the asymmetry becomes less pronounced. Details about the convergence of our algorithm are given in Sec.~1.3 of the SM~\cite{supplement}.

\section{ Conclusion}
This work presents a powerful optimization tool for adiabatic control of open few-body quantum systems. 
Generalizing~\cite{brif2014exploring}, we seek pulses under which the evolving system adheres  to the instantaneous eigenstate along its  propagation. Combining efficient optimization and simulation tools, we can optimize technologically relevant systems, as demonstrated by the three numerical examples above. 
We conclude by mentioning directions for further research. Adherence to the instantaneous ground state along the propagation requires knowledge of the state, which is not available in large-scale AQC. Nevertheless, we hypothesize that AQC can be improved  by seeking pulses that adhere to approximate ground states or by using knowledge of the ground state  during early  stages of the protocol. Furthermore, optimization of AQC requires classical computation resources in the optimization stage and quantum resources in the evolution stage. Optimizing  the division between classical and quantum resources  in adiabatic QOC is key to advancing the present  approach.

\emph{Acknowledgment: }
AP acknowledges support from the Israel Science Foundation  Grant No. 1484/24 and the   Alon Fellowship of the Israeli Council of Higher Education.
The authors thank Prof. Nir Davidson, Prof. Ofer Firstenbeg, and  Amit Tsabari for insightful discussions on noise sources in adiabatic control of atoms. The authors thank Dr. Haggai Landa for discussions on noise sources in the IBM quantum platform. 
Finally, the authors thank  Nadav Avtar and Dor Atias for support with the digitized RAP code.

\clearpage
\widetext

\setcounter{section}{0}
\pagenumbering{gobble}

\setcounter{figure}{0}
\setcounter{table}{0}
\renewcommand{\thefigure}{S\arabic{figure}}
\renewcommand{\thetable}{S\arabic{table}}
\addto\captionsenglish{\renewcommand{\figurename}{Fig.}}

\section*{Supplementary Information for Gradient-Free Pulse Optimization for Adiabatic Control
in Open Few-Body Quantum Systems}

This supplementary information contains details about the optimization tools employed in the main text. The first section (\secref{overview}) introduces the algorithm used in the main text. called adiabatic QOC, which is based on CMA-ES. Subsections discuss the bases chosen for pulse descretization, the weights of cost terms in multi-objective optimization, and analysis of convergence of our method. Then, sections 2--4 present complementary details to topics introduced in the main text.



\tableofcontents
\section{Algorithm overview: Adiabatic QOC with CMA-ES \label{sec:overview}}
In this section, we briefly summarize the covariance matrix adaptation evolution strategy (CMA-ES). 
Unlike gradient-based or deterministic optimization methods such as Krotov, GRAPE, or Nelder--Mead, 
CMA-ES is a stochastic, population-based approach. 
At each iteration, a population of candidate solutions is sampled from a multivariate normal distribution in the search space. 
This distribution is defined by three adaptive parameters: the mean  $\mu$, a global step-size $\sigma$, and a covariance matrix $C$. 
The mean directs the search toward promising regions, the step-size determines the overall exploration scale, 
and the covariance matrix captures correlations between parameters, thereby identifying effective search directions. 
In the context of optimal control, the search space corresponds to the coefficients of the chosen basis parametrization.

\begin{figure}[t]  \textbf{Pseudocode for Adiabatic QOC Algorithm using  CMA-ES }
\begin{algorithmic}[1]
\Statex \textbf{Input:}
\Statex \hspace{\algorithmicindent} $\{f_m(t)\}$: Expansion basis
\Statex \hspace{\algorithmicindent} $M$: Number of basis functions
\Statex \hspace{\algorithmicindent} $N_{\text{iter}}$: Number of iterations
\Statex \hspace{\algorithmicindent} 
{$N_{\text{samp}}$: Number of  samples 
\Statex \hspace{\algorithmicindent} }
$f_0(t)$: Reference pulse
\For{$k = 1,\dots,N_{\text{iter}}$}
\State Pick $N_{\text{samp}}$ samples of  weights $w_m^{(k,i)}$ 
       \Comment{$k$: iteration, $i$: sample, $m$: basis function}
\Statex \hspace{\algorithmicindent}\Comment{Weights drawn from $\mathcal{N}(\mu_m^{(k)},\sigma^{(k)})\quad\quad\quad\,$}
  
  \For{each coefficient sample $i = 1,\dots, N_\mathrm{samp}$} \Comment{Adiabatic QOC loop}
    \State Construct pulse, $v^{(k,i)}(t) = f_0(t) + \sum_{m=1}^{M} w_m^{(k,i)} f_m(t)$. \Comment{Pulse construction$\quad$}
    \State Evolve Lindblad equation [Eq. (1) from main text]. \Comment{Evolution$\,\,\,\,\,\quad\quad\quad\quad$}
    \State Compute cost function, $C^{(k,i)}$ [Eq. (2) in main text] \Comment{Evaluation$\quad\quad\quad\quad\,$}. 
  \EndFor
  \State Select best candidates and update distribution $\mathcal{N}(\mu_m^{(k+1)},\sigma^{(k+1)})$ and covariance $\Sigma^{(k+1)}$~\cite{bennet2021nevergrad,hansen2006cma}. 
\EndFor
\State \textbf{Output:} Optimal coefficients $w_m^{(*)}$ and optimized pulse: $v^*(t) = f_0(t) + \sum_{m=1}^{M} w^*_m f_m(t)$.
\end{algorithmic}
\end{figure}

\subsection{Weights of terms in the  cost function}

In the main text, we compare the impact of the cost functions $C_1$ and $C_2$ on the optimization results. 
The $C_2$ objective function minimizes the deviation from the desired instantaneous eigenstate throughout the evolution, 
thereby enhancing robustness against certain types of control errors. 
However, this improvement in robustness comes at the expense of a slightly reduced final-state fidelity.

\begin{figure}[H]
    \centering
    \includegraphics[width=1\linewidth]{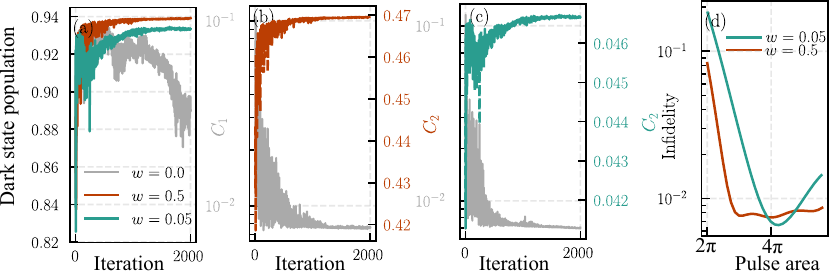}
    \caption{(a) Average dark state occupation during the optimization process for different weights of the $C_2$ term:
    $w = 0$ (gray), $w = 0.5$ (red), and $w = 0.05$ (cyan). 
    (b,c) Convergence of the infidelity for and $\lambda C_2$ throughout the optimization process for $\lambda =0.5,0.05$. 
     (d) infidelity vs pulse area }
    \label{fig:placeholder}
\end{figure}

\subsection{Comparing  discretization bases}
 In the main text, we use  three bases [Eq. (5)] to parametrize the control pulses in the CRAB algorithm: sinusoidal functions in \emph{example I} (RAP), Chebyshev polynomials in \emph{example II} (multimode STIRAP), and Gaussian functions in \emph{example III} (MIS). 
In the Gaussian basis, we construct pulses using sums of   7 Gaussians [Eq.~(4a)] whose amplitude, widths, and centers are sampled uniformly form the range: $w_m \in (\Omega_0/1000, \Omega_0/10)$, 
 $\sigma_m \in (0.05,\; 0.6)$, and $\mu_m \in(0.06\,t_f/2,\; 1.8\,t_f/2)$
respectively, with $\Omega_0$ the maximum Rabi frequency of the reference pulse and $t_f$ the protocol duration. In the cases of Chebyshev and sinusoidal bases [Eqs.~(4b,c)], the expansion weights, $w_m$, are sampled from a normal distribution with zero mean and unit standard deviation.

\begin{figure}[t!]
    \centering
    \includegraphics[width=0.85\linewidth]{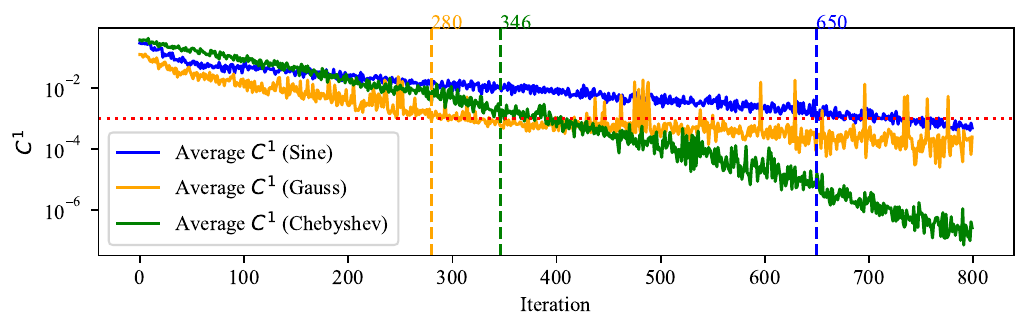}
    \caption{Average cost $C_1$ as a function of optimization iteration, when optimizing RAP. We compare the performance using three bases:    Sinusoidal functions (blue), Gaussians (orange), and Chebyshev polynomials (green)  and sampling from  50 random seeds in the CMA-ES algorithm.}
    \label{fig:basis_comparison}
\end{figure}
Next, we justify our basis choice by analyzing  the convergence rate and optimal performance of  each of the bases. We compare the performance of the three bases  in  rapid adiabatic  passage (RAP).  
Our comparative analysis shows that  the suitability of a given basis depends  on the optimization objective as follows:  When the primary goal is to maximize  the final-time fidelity [$C_1$ in Eq.~(3a)], Chebyshev polynomials consistently achieve lower minima 
of the cost function. In contrast, when seeking a  rapid convergence rate, the Gaussian basis yields more favorable results. The sinusoidal basis  achieved a compromise of final state fidelity and convergence rate.

To illustrate this point, we fix the detuning profile of the RAP pulses  to $\Delta(t)=\Delta_0\left(\frac{2t}{Tf}-1\right)$ and search for optimal shape of the Rabi frequency  $\Omega(t)$ using the three above-mentioned bases. For each basis, we choose six basis functions and run the CMA-ES algorithm to optimize $C^1_\mathrm{fid}$. Recall that the algorithm consists of optimizing the figure of merit $C^1_\mathrm{fid}$ over the control parameter space for an ensemble of random choices of   controls, where in each iteration, the average and standard deviation of the set is computed and used to generate a new sample (as detailed in Sec.~4). Here, we average over 50 samples.


Benchmarking revealed that Chebyshev polynomials are best suited 
for the multimode-STIRAP protocol, where achieving very high fidelities is critical, while Gaussian functions 
were preferred for the MIS problem since the convergence rate is a central issue in optimizing multi-qubit systems.

\subsection{Analyzing convergence of CMA-ES}

In the main text, we apply adiabatic quantum optimal control (QOC) to solve the Maximum Independent Set (MIS) problem on ring graphs of length $N = 2$ to $14$. Here, we analyze the convergence behavior of the optimization algorithm for three  instances with $N = 8$, $10$, and $12$ atoms. To monitor optimization progress, we track the fidelity at each iteration. The control pulse is parameterized using a sum of seven Gaussian basis functions see Eq.~5c in the main text, with each Gaussian defined by an amplitude, width, and temporal offset, yielding a 21-dimensional optimization space. 

\begin{figure}[H]
    \centering
    \includegraphics[width=\linewidth]{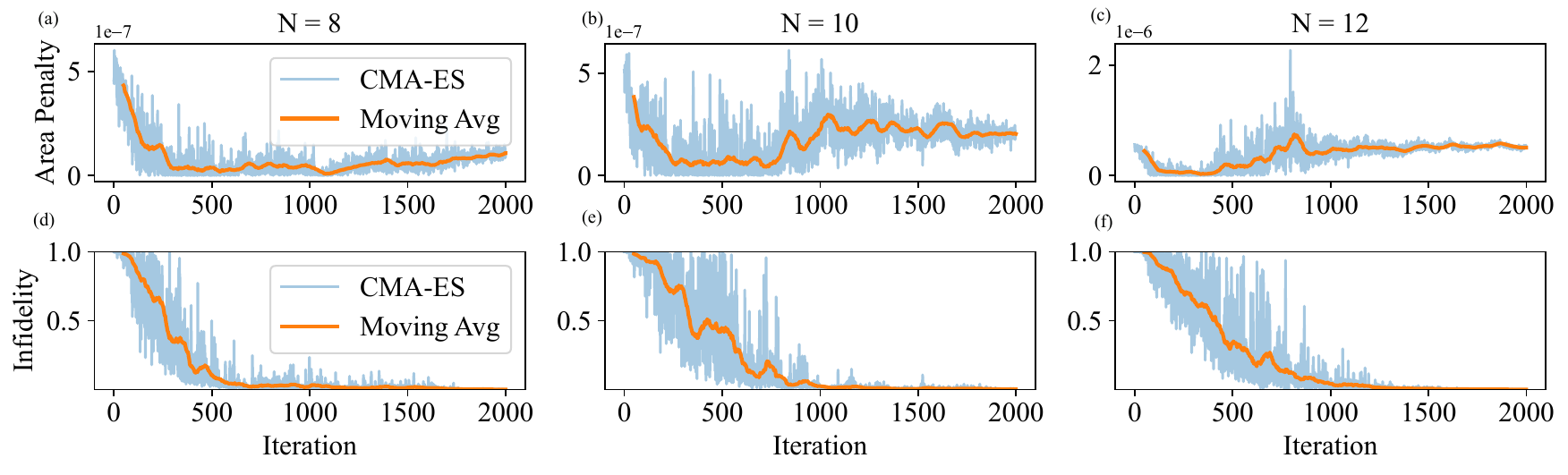}
    \caption{
    Convergence behavior of the optimization process for $N = 8$, $10$, and $12$ atoms. 
    Panels (a)–(c) show the evolution of the area penalty over 2000 iterations, while panels (d)–(f) display the corresponding infidelity. 
    Both raw values and moving averages (window size 50) are shown. For all instances, the optimization exhibits improvement, with final fidelities approaching unity.
    }
    \label{convergence}
\end{figure}

Figure~\ref{convergence} shows  the cost function terms as a function of iteration number of rings containing $N = 8, 10$ and 12 nodes. The top row displays both raw and moving-averaged traces of the area penalty, and the bottom row shows the fidelity on a linear scale. Across all three values of $N$, we observe two characteristic thresholds in convergence: an initial point where the infidelity begins to drop significantly, and a later stage where it stabilizes near its final value. Both thresholds shift to later iterations for larger $N$, indicating increased complexity in the optimization landscape. 

In the main text, the area of the pulses obtained from the optimization algorithm is renormalized before running the AQC. We do so in order to compare the performance of pulses that have an equal pulse area. The infidelity values shown in Fig.~(S2) (bottom) correspond to the pulses before renormalizing their area. This explains the slight deviation between the infidelity values shown here and in the main text.

\section{Robustness of adiabatic QOC}

\subsection{Benchmarking adiabatic QOC and ensemble optimization in RAP}

In this work, we compare the performance of adiabatic QOC (see main text) and ensemble QOC. Adiabatic QOC directly minimizes the time-integrated diabatic error, $(C_2)$,[Eq.~(3, b)], steering the system to follow the instantaneous ground state throughout the evolution. By contrast, ensemble optimization minimizes the weighted-average terminal diabatic error, $(C_3)$ [Eq.~(3, c), \cite{goerz2014robustness}], across an ensemble of Hamiltonians with perturbed amplitudes, detunings, or lifetimes to ensure robustness against parameter fluctuations. The principal benefit of adiabatic QOC is its dramatic reduction in classical optimization runtime while achieving comparable robustness.

Physically, spatial inhomogeneities in our control beams and residual atomic motion give rise to two dominant error channels in RAP implementations: variations in the integrated pulse area 
and Doppler‐induced detuning offsets. 
An atom displaced off-axis experiences a reduced Rabi frequency, while thermal velocities detune the drive. Demonstrating that adiabatic QOC pulses maintain high fidelity across these amplitude and frequency fluctuations confirms that our optimization protocol is resilient to the primary noise sources encountered in realistic laboratory quantum-control experiments.

A detailed comparison of the robustness landscapes obtained via adiabatic QOC [Eq.(3, b)] and ensemble optimization [Eq.(3, c)] is presented in \figref{ SUP_1}. In Python simulations (\figref{ SUP_1},a--b), both schemes yield nearly identical high-fidelity regions across the pulse-area–Doppler-shift plane, with infidelities remaining below $10^{-3}$ over the central plateau. Trotterized experiments on IBM hardware (\figref{ SUP_1},c--d) reproduce these trends, confirming that the optimized pulses perform equally well in the laboratory. Although the robustness landscapes are nearly identical, the classical runtime differs dramatically: all CMA-ES optimizations were run single-threaded on an ASUS laptop with an Intel Core Ultra 9 185H processor (16 cores, 22 threads @ 2.50 GHz base, 32 GB RAM) under Windows 11 Home 24H2. Using a fixed budget of 2000 CMA-ES iterations, adiabatic QOC completed in \(\sim\!2\) min, whereas ensemble optimization—evaluated over a \(5\times5\) grid of pulse areas and Doppler shifts at the same budget—required \(\sim\!40\) min, underscoring the superior computational efficiency of adiabatic QOC

To verify the robustness of our optimized pulses on real quantum hardware, we digitized the continuous RAP Hamiltonian using a second-order Trotter decomposition~\cite{Suzuki1976} and executed the resulting circuits on IBM Quantum.  
In our Qiskit implementation, the instantaneous unitary
\begin{equation}
U(t_k + \Delta t, t_k) \;=\; \exp\bigl(-i\,H(t_k)\,\Delta t\bigr),
\quad
H(t_k)=\Omega(t_k)\,\sigma_x+\Delta(t_k)\,\sigma_z,
\quad
\Delta t = t_{k+1}-t_k.
\label{eq:trotter_unitary}
\end{equation}
was approximated via the symmetric second-order Suzuki–Trotter decomposition~\cite{Suzuki1976}:
\begin{equation}
\exp\bigl(-i\,H(t_k)\,\Delta t\bigr)
\;\approx\;
R_z\!\bigl(\Delta(t_k)\,\Delta t/2\bigr)\,
R_x\!\bigl(\Omega(t_k)\,\Delta t\bigr)\,
R_z\!\bigl(\Delta(t_k)\,\Delta t/2\bigr).
\label{eq:second-order_Suzuki-Trotter_decomposition}
\end{equation}
Here, \(R_x(\theta)=e^{-i\,\theta\,\sigma_x/2}\) and 
\(R_z(\phi)=e^{-i\,\phi\,\sigma_z/2}\) denote single-qubit rotations 
about the \(x\)- and \(z\)-axes, respectively.
implemented via Qiskit’s \texttt{rz} and \texttt{rx} gates in the \texttt{trot\_circuit} routine.

All 625 parameter-sweep circuits were transpiled for the IBM Quantum backend \texttt{ibm\_brisbane} 
accessed through \texttt{QiskitRuntimeService}~\cite{IBMQuantum}.  For consistency across runs, each single-qubit circuit was pinned to physical qubit 3, with optimization level 0 and measurement in the Z-basis.  The resulting infidelity landscape (\figref{ SUP_1}c–d) closely matches our  Python simulations, confirming that the digitized adiabatic pulses retain their designed robustness under the native gate set and noise characteristics of real hardware.

\begin{figure}[H]
    \centering
    \includegraphics[width=1\linewidth]{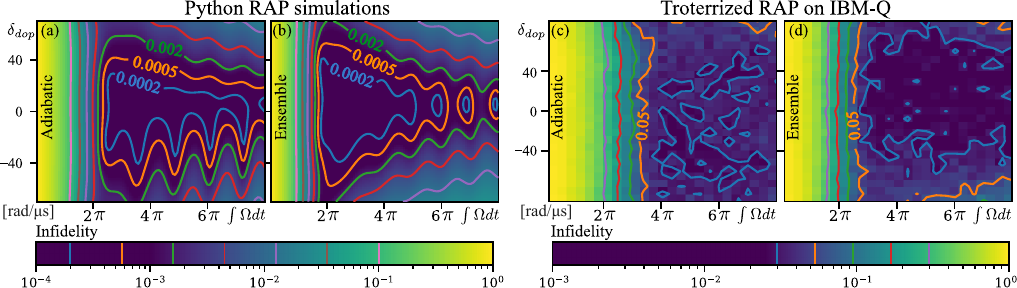}
    \caption{Robustness of RAP pulses. (a,b) Python simulations showing infidelity of state transfer fidelity (see colorbar), as a function of pulse area and Doppler shift for pulses optimized via adiabatic QOC [Eq.(3, b)] (a) and ensemble optimization [Eq.(3, c)] (b). (c,d) Trotterized RAP experiments on IBM hardware~\cite{IBMQuantum} for the same two optimization schemes: adiabatic QOC (c) and ensemble (d).}

\label{fig: SUP_1} 
\end{figure}

\subsection{Robustness of  optimized pulses in the MIS AQC}

\begin{figure}[t]
\noindent
\begin{minipage}{0.25\textwidth} 
    \includegraphics[width=\linewidth]{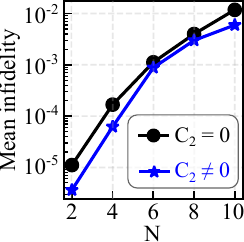}
\end{minipage}%
\hspace{0.03\textwidth}%
\begin{minipage}{0.5\textwidth} 
    \raggedleft 
    \caption{Mean infidelity under a variation of $\pm 10\%$ of the optimal Rabi amplitude, 
        comparing results obtained with $C_2$ included in the quantum optimal control 
        optimization (blue stars) and without $C_2$ (black dots), versus the number of qubits.}
    \label{fig:MIS_ROBUST}
\end{minipage}
\end{figure}

A source of error in atomic quantum computation arises from fluctuations in the positions of atoms confined in optical tweezers. 
To evaluate the performance of the cost functions $C_1$ and $C_2$, we tested their robustness against these fluctuations as a deviation from the ideal laser amplitude as demonstrated in \figref{MIS_ROBUST}(b)

Due to the competition between adiabatic evolution and decoherence processes, inclusion of $C_2$ may decrease the transfer fidelity at high dephasing rates. The optimal weighting of $C_2$ should therefore be chosen with respect to the specific process and the characteristic timescale of the experiment.\figref{MIS_ROBUST}(a)

\section{Hilbert-space truncation  error in Rydberg simulations}

We now analyze the validity of the independent set subspace approximation.  
The dynamics are governed by the Rydberg Hamiltonian
\begin{equation}
    H = \Omega(t)\sum_i \hat\sigma_x^i 
    - \Delta(t)\sum_i \frac{I - \hat\sigma_z^i}{2} 
    + V \sum_{ij} \hat{n_i} \hat{n_j} ,
\end{equation}
where $\Omega(t)$ is the Rabi frequency, $\Delta(t)$ the detuning, and $V$ the interaction strength between atoms in Rydberg states.  In the blockade limit $V \rightarrow \infty$, simultaneous excitations on adjacent sites are strongly suppressed, as one atom in the Rydberg state detunes its neighbors from resonance. The accessible Hilbert space is then restricted to the independent-set subspace, which can be enforced through the projector $\mathcal{P}=\underset{ij\in E}{\sum}n_in_j$
\begin{equation}
   \mathcal{P}\KET{\psi} = 0 ,
\end{equation}

To assess the quality of this truncation, we compare the full Hilbert space dynamics and reduced independent set dynamics by evaluating the overlap
\begin{equation}
    \epsilon = \big|\langle{\psi_{\mathrm{reduced}}(T_f)|\psi_{\mathrm{full}}(T_f)}\rangle\big|^2 ,
\end{equation}
where $\KET{\psi_{\mathrm{reduced}}(T_f)}$ and $\KET{\psi_{\mathrm{full}}(T_f)}$ denote the reduced subspace and full space wave functions, respectively, at the final evolution time $T_f$. Here $\epsilon$ serves as a fidelity measure, with deviations from unity quantifying the truncation error.  

In addition, we evaluate the leakage into doubly excited states by monitoring the expectation value of $\sum_{ij} n_i n_j$ at $T_f$. This provides a direct measure of blockade violation in the full dynamics.  

\begin{figure}[h]
    \centering
    \includegraphics[width=1\linewidth]{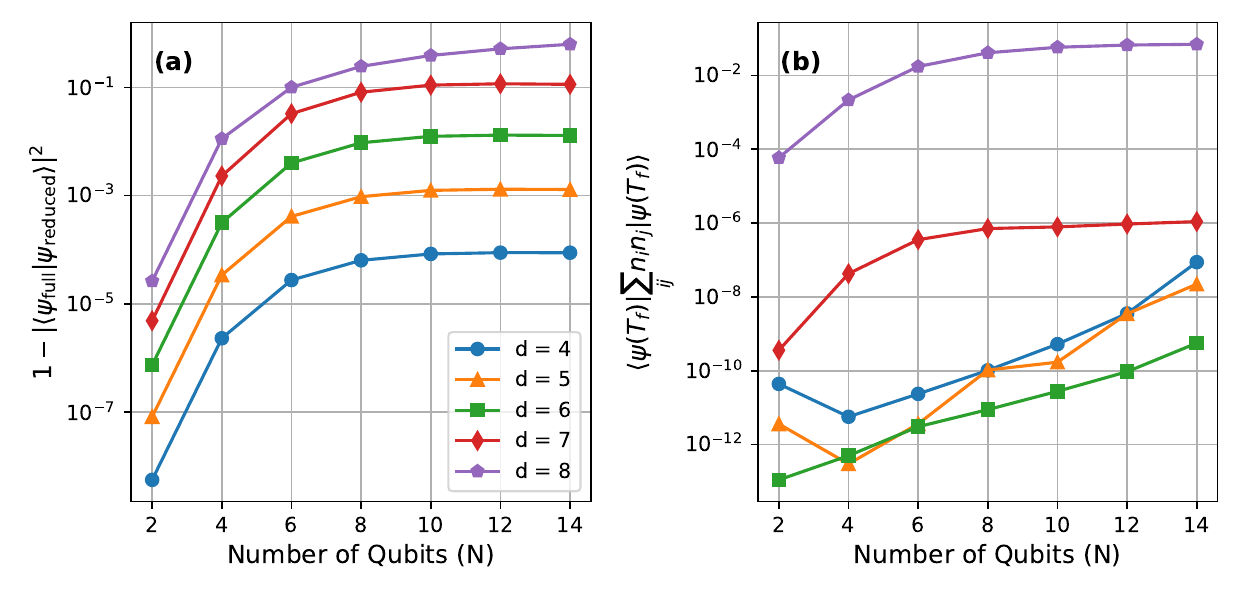}
    \caption{(a) Truncation error $\epsilon = \big|\langle\psi_{\mathrm{reduced}}(T_f)|\psi_{\mathrm{full}}(T_f)\rangle\big|^2 $ as a function of system size $N$ for several adjacent atoms  distance $d$ in $\mu \mathrm{m}$. (b) Expectation value of the doubly excited population $\langle \sum_{i,j} n_i n_j \rangle$ at the final time $T_f$ as a function of $N$.} 
\end{figure}

\section{Super-adiabatic transitionless driving for qubit state transfer}

The system from Ref.~\cite{chang2020remote}  consists of two  qubits, which are coupled to a long multimode waveguide via tunable couplers  [Fig.~3(a) in the main text].
Qubits $A$ and $B$ are coupled to  the $n^{th}$  mode  of the waveguide with amplitudes $g_{ac}^n(t)$ and $g_{bc}^n(t)$ respectively.
Considering  the single-excitation subspace (i.e., neglecting multiphoton excitations in the connector),     the system's Hamiltonian is~\cite{malekakhlagh2024enhanced,schwartzman2025modeling}:
\begin{gather} 
    H_{\mathrm{STIRAP}} = \scalebox{0.95}{$
    \sum_n \left\{ 
    \Delta_n |0_a0_b1_{c}^{(n)}\rangle \langle0_a0_b1_{c}^{(n)}|+   \right.$} 
   \scalebox{0.95}{$\left.
g_{ac}^n(t)\KET{1_a0_b0_{c}}\langle{0_a0_b1_{c}^{(n)}|}   + g_{bc}^n(t)\KET{0_a1_b0_{c}}\langle{0_a0_b1_{c}^{(n)}}| +\rm{h.c.}\right\}$}.
\label{eq:H-lambda}
\end{gather}
The notation $\KET{1_a0_b0_{c}}$ indicates  that qubits $A$ and $B$ are  in  $\KET{1}$ and  $\KET{0}$ respectively,  and all guided modes have zero photons.  Similarly,  $|0_a0_b1_{c}^{(n)}\rangle$ indicates  that both qubits are in  $\KET{0}$, the $n^{th}$ guided mode contains a single photon, while  all other  modes are empty. We use $\Delta_n = n\Delta_c$ to denote the  detuning of the coupling to the $n^{th}$ guided mode, with $n \in \mathbb{Z}$ and  $\Delta_c$ denoting the free spectral range (FSR) of the waveguide.  We assume that the coupling coefficients $g_{ab}^{(n)}, g_{ac}^{(n)}$ are equal in magnitude  for all $n$ and  have alternating signs [$\propto(-1)^n$] when the parity of guided modes alternates between even and odd.
 
When $\Delta_c\gg g_{ab}^{(n)}, g_{ac}^{(n)}$, only  the resonant $n = 0$ mode  couples to the qubits and the single-excitation subspace contains three levels, as in traditional STIRAP. The  Hamiltonian has a ``dark state,'' containing zero photons, 
\begin{equation}
\KET{D(t)} =\cos\theta(t)\KET{1_a0_b0_c} -\sin\theta(t) \KET{0_a1_b0_c},
\end{equation}
with $\theta(t) \equiv \tan^{-1} [g_{ac}(t)/g_{bc}(t)]$ .  
The dark state changes from $\KET{1_a0_b0_c}$   to $\KET{0_a1_b0_c}$
 by changing $\theta(t)$   from $0$ to $\pi/2$, i.e.,  by applying first $g_{bc}(t)$ and then $g_{ac}(t)$ [Fig.~3(b)]. STIRAP relies on adiabatically following 
 $\KET{D(t)}$.
 To accelerate  STIRAP, one can use that SATD pulses~\cite{malekakhlagh2024enhanced}
\begin{gather} 
g_{ac}^{\rm{SATD}}(t)  = g \left(\sin\theta(t) + \frac{\cos[\theta(t)]\ddot{\theta}(t)}{g^2+\dot{\theta}(t)^2}\right)\nonumber\\
g_{bc}^{\rm{SATD}}(t) = g\left(\cos\theta(t) - \frac{\sin[\theta(t)]\ddot{\theta}(t)}{g^2+\dot{\theta}(t)^2} \right).
\label{eq:SATD-fields}
\end{gather} 
In the limit  of large FSR,  SATD pulses  give excellent performance since they are obtained by analytically eliminating diabatic errors.

However, when the FSR is comparable to the coupling amplitudes $\Delta_c \simeq \max(g_{ab}^{(n)}, g_{ac}^{(n)})$, the SATD solution is not exact. The next-order  approximation to the dynamics includes  the  nearest sideband modes ($n\pm1$),  and the single excitation subspace contains  five levels. The Hamiltonian has a quasi dark state~\cite{malekakhlagh2024enhanced}
\begin{gather} 
|{D^{(1)}(t)}\rangle  \propto 
\cos(\theta)\,\KET{1_a0_b0_c} 
-  \sin(\theta)\KET{0_a1_b0_c} +
\tfrac{\sin(2\theta)g(t)}{\Delta}|0_a0_b1_c^{(-1)}\rangle  
+ \tfrac{\sin(2\theta)g(t)}{\Delta}|0_a0_b1_c^{(1)} \rangle. 
\label{eq:quasi-dark}
\end{gather}
The second line contains the bright-state (single-photon) wavefunction components, which are  suppressed by increasing the ratio $|g(t)|/\Delta_c$, where $g\equiv\sqrt{g_{ac}^2+g_{bc}^2}$. A pulse obtained with adiabatic QOC, i.e. optimizing the adherence to $|{D^{(1)}(t)}\rangle $ using a basis of Chebyshev polynomials [Eq.~(4c)] is shown in Fig.~3(b) (solid).

\endwidetext

%

\end{document}